\documentclass[sigconf]{acmart}

\usepackage[T1]{fontenc}
\usepackage{enumitem}
\usepackage{afterpage}
\usepackage{listings}
\usepackage{xcolor}
\usepackage{graphicx}
\usepackage{float}
\usepackage{caption}
\usepackage{subcaption}
\usepackage{placeins}
\usepackage{url}
\usepackage{epstopdf}
\usepackage{booktabs}
\usepackage{footnote}
\makesavenoteenv{table}
\makesavenoteenv{tabular}

\usepackage[acronyms,nonumberlist,nopostdot,nomain,nogroupskip]{glossaries}

\usepackage{tabularx}

\usepackage{paralist}

\newacronym{quic}{QUIC}{Quick UDP Internet Connections}
\newacronym{3gpp}{3GPP}{3rd Generation Partnership Project}
\newacronym{adc}{ADC}{Analog to Digital Converter}
\newacronym{5g}{5G}{5th generation}
\newacronym{aimd}{AIMD}{Additive Increase Multiplicative Decrease}
\newacronym{am}{AM}{Acknowledged Mode}
\newacronym{amc}{AMC}{Adaptive Modulation and Coding}
\newacronym{aqm}{AQM}{Active Queue Management}
\newacronym{awgn}{AGWN}{Additive White Gaussian Noise}
\newacronym{balia}{BALIA}{Balanced Link Adaptation}
\newacronym{bdp}{BDP}{Bandwidth-Delay Product}
\newacronym{bf}{BF}{Beamforming}
\newacronym{cc}{CC}{Congestion Control}
\newacronym{cdf}{CDF}{Cumulative Distribution Function}
\newacronym{cn}{CN}{Core Network}
\newacronym{cqi}{CQI}{Channel Quality Information}
\newacronym{cp}{CP}{Control Plane}
\newacronym{csirs}{CSI-RS}{Channel State Information - Reference Signal}
\newacronym{dc}{DC}{Dual Connectivity}
\newacronym{dce}{DCE}{Direct Code Execution}
\newacronym{dci}{DCI}{Downlink Control Information}
\newacronym{dl}{DL}{Downlink}
\newacronym{dmr}{DMR}{Deadline Miss Ratio}
\newacronym{dmrs}{DMRS}{DeModulation Reference Signal}
\newacronym{e2e}{E2E}{End-to-End}
\newacronym{ecn}{ECN}{Explicit Congestion Notification}
\newacronym{edf}{EDF}{Earliest Deadline First}
\newacronym{enb}{eNB}{evolved Node Base}
\newacronym{epc}{EPC}{Evolved Packet Core}
\newacronym{es}{ES}{Edge Server}
\newacronym{fdma}{FDMA}{Frequency Division Multiple Access}
\newacronym{fdd}{FDD}{Frequency Division Duplexing}
\newacronym[firstplural=Radio Access Technologies (RATs)]{rat}{RAT}{Radio Access Technology}
\newacronym{fs}{FS}{Fast Switching}
\newacronym{ftp}{FTP}{File Transfer Protocol}
\newacronym{gnb}{gNB}{Next Generation Node Base}
\newacronym{harq}{HARQ}{Hybrid Automatic Repeat reQuest}
\newacronym{hetnet}{HetNet}{Heterogeneous Network}
\newacronym{hh}{HH}{Hard Handover}
\newacronym{hol}{HOL}{Head-of-Line}
\newacronym{ia}{IA}{Initial Access}
\newacronym{imt}{IMT}{International Mobile Telecommunication}
\newacronym{iot}{IoT}{Internet of Things}
\newacronym{los}{LOS}{Line of Sight}
\newacronym{lte}{LTE}{Long Term Evolution}
\newacronym{m2m}{M2M}{Machine to Machine}
\newacronym{mac}{MAC}{Medium Access Control}
\newacronym{mc}{MC}{Multi-Connectivity}
\newacronym{mcs}{MCS}{Modulation and Coding Scheme}
\newacronym{mec}{MEC}{Mobile Edge Cloud}
\newacronym{mi}{MI}{Mutual Information}
\newacronym{mimo}{MIMO}{Multiple Input, Multiple Output}
\newacronym{mmwave}{mmWave}{millimeter wave}
\newacronym{mr}{MR}{Maximum Rate}
\newacronym{mss}{MSS}{Maximum Segment Size}
\newacronym{mtd}{MTD}{Machine-Type Device}
\newacronym{mtu}{MTU}{Maximum Transmission Unit}
\newacronym{nfv}{NFV}{Network Function Virtualization}
\newacronym{nlos}{NLOS}{Non Line of Sight}
\newacronym{nr}{NR}{New Radio}
\newacronym{ofdm}{OFDM}{Orthogonal Frequency Division Multiplexing}
\newacronym{pdcch}{PDCCH}{Physical Downlonk Control Channel}
\newacronym{pdcp}{PDCP}{Packet Data Convergence Protocol}
\newacronym{pdsch}{PDSCH}{Physical Downlink Shared Channel}
\newacronym{pdu}{PDU}{Packet Data Unit}
\newacronym{pf}{PF}{Proportional Fair}
\newacronym{pgw}{PGW}{Packet Gateway}
\newacronym{phy}{PHY}{Physical}
\newacronym{pbch}{PBCH}{Physical Broadcast Channel}
\newacronym[plural=\gls{mme}s,firstplural=Mobility Management Entities (MMEs)]{mme}{MME}{Mobility Management Entity}
\newacronym{prb}{PRB}{Physical Resource Block}
\newacronym{pss}{PSS}{Primary Synchronization Signal}
\newacronym{pucch}{PUCCH}{Physical Uplink Control Channel}
\newacronym{pusch}{PUSCH}{Physical Uplink Shared Channel}
\newacronym{rach}{RACH}{Random Access Channel}
\newacronym{ran}{RAN}{Radio Access Network}
\newacronym{red}{RED}{Random Early Detection}
\newacronym{rf}{RF}{Radio Frequency}
\newacronym{rlc}{RLC}{Radio Link Control}
\newacronym{rlf}{RLF}{Radio Link Failure}
\newacronym{rrc}{RRC}{Radio Resource Control}
\newacronym{rrm}{RRM}{Radio Resource Management}
\newacronym{rr}{RR}{Round Robin}
\newacronym{rs}{RS}{Remote Server}
\newacronym{rsrp}{RSRP}{Reference Signal Received Power}
\newacronym{rss}{RSS}{Received Signal Strength}
\newacronym{rtt}{RTT}{Round Trip Time}
\newacronym{rw}{RW}{Receive Window}
\newacronym{rx}{RX}{Receiver}
\newacronym{sa}{SA}{standalone}
\newacronym{sack}{SACK}{Selective Acknowledgment}
\newacronym{sap}{SAP}{Service Access Point}
\newacronym{sch}{SCH}{Secondary Cell Handover}
\newacronym{scoot}{SCOOT}{Split Cycle Offset Optimization Technique}
\newacronym{sdma}{SDMA}{Spatial Division Multiple Access}
\newacronym{sinr}{SINR}{Signal to Interference plus Noise Ratio}
\newacronym{sm}{SM}{Saturation Mode}
\newacronym{snr}{SNR}{Signal to Noise Ratio}
\newacronym{son}{SON}{Self-Organizing Network}
\newacronym{ss}{SS}{Synchronization Signal}
\newacronym{srs}{SRS}{Sounding Reference Signal}
\newacronym{sss}{SSS}{Secondary Synchronization Signal}
\newacronym{tb}{TB}{Transport Block}
\newacronym{tcp}{TCP}{Transmission Control Protocol}
\newacronym{tdd}{TDD}{Time Division Duplexing}
\newacronym{tdma}{TDMA}{Time Division Multiple Access}
\newacronym{tfl}{TfL}{Transport for London}
\newacronym{tm}{TM}{Transparent Mode}
\newacronym{trp}{TRP}{Transmitter Receiver Pair}
\newacronym{tti}{TTI}{Transmission Time Interval}
\newacronym{ttt}{TTT}{Time-to-Trigger}
\newacronym{tx}{TX}{Transmitter}
\newacronym{ue}{UE}{User Equipment}
\newacronym{ul}{UL}{Uplink}
\newacronym{uml}{UML}{Unified Modeling Language}
\newacronym{um}{UM}{Unacknowledged Mode}
\newacronym{utc}{UTC}{Urban Traffic Control}
\newacronym{vm}{VM}{Virtual Machine}
\newacronym{rsrq}{RSRQ}{Reference Signal Received Quality}
\newacronym{rssi}{RSSI}{Received Signal Strength Indicator}
\newacronym{crs}{CRS}{Cell Reference Signal}
\newacronym{comp}{CoMP}{Coordinated Multi-Point}
\newacronym{cran}{C-RAN}{Cloud \acrlong{ran}}
\newacronym{ca}{CA}{Carrier Aggregation}
\newacronym{cco}{CC}{Carrier Component}
\newacronym{nsa}{NSA}{Non Stand Alone}
\newacronym{embb}{eMBB}{Enhanced Mobility Broadband}
\newacronym{bsr}{BSR}{Buffer Status Report}
\newacronym{srb}{SRB}{Service Radio Bearer}
\newacronym{scm}{SCM}{Spatial Channel Model}
\newacronym{sctp}{SCTP}{Stream Control Transmission Protocol}
\newacronym{mptcp}{MPTCP}{Multi-path TCP}
\newacronym{ietf}{IETF}{Internet Engineering Task Force}
\newacronym{os}{OS}{Operating System}
\newacronym{tls}{TLS}{Transport Layer Security}
\newacronym{rfc}{RFC}{Request for Comments}
\newacronym{http}{HTTP}{HyperText Transfer Protocol}
\newacronym{nat}{NAT}{Network Address Translation}
\newacronym{api}{API}{Application Programming Interface}
\newacronym{rto}{RTO}{Retransmission Timeout}
\newacronym{psc}{PSC}{Public Safety Communication}
\newacronym{rpgm}{RPGM}{Reference Point Group Mobility}
\newacronym{ic}{IC}{Incident Command}
\newacronym{rsu}{RSU}{Road Side Unit}
\newacronym{uav}{UAV}{Unmanned Aerial Vehicle}
\newacronym{iab}{IAB}{Integrated Access and Backhaul}

\usepackage{tikz}
\usepackage{pgfplots}
\pgfplotsset{compat=newest} 
\pgfplotsset{plot coordinates/math parser=false} 
\newlength\fheight
\newlength\fwidth
\usetikzlibrary{plotmarks,patterns,decorations.pathreplacing,backgrounds,calc,arrows,arrows.meta,spy,matrix}
\usepgfplotslibrary{patchplots,groupplots}
\usepackage{tikzscale}

\begin{document}


\flushbottom
\setlength{\parskip}{0ex plus0.1ex}
\addtolength{\skip\footins}{-0.2pc plus 40pt}

\title{Implementation of Reference Public Safety Scenarios in ns-3}




\author{\texorpdfstring{Michele Polese, Tommaso Zugno, Michele Zorzi\\
\small Department of Information Engineering, University of Padova, Padova, Italy \\
\small e-mail: \{polesemi, zugnotom, zorzi\}@dei.unipd.it}{}}

\copyrightyear{2019}
\acmYear{2019}
\setcopyright{none}
\acmConference[WNS3 '19]{2018 Workshop on ns-3}{June 19--20, 2019}{Firenze, Italy}
\acmBooktitle{WNS3 '19: 2019 Workshop on ns-3, June 19--20, 2019, Firenze, Italy}
\acmPrice{xxxx}
\acmDOI{xxxxxxxxxxxxxxxxx}
\acmISBN{xxxxxxxxxxxxxxxx}

\pagestyle{empty}

\begin{abstract}

During incidents and disasters it is fundamental to provide to first responders high performance and reliable communications, in order to improve their coordination capabilities and their awareness of the surrounding environment, and to allow them to promptly transmit and receive alerts on possible dangerous situations or emergencies. The accurate evaluation of the performance of different \gls{psc} networking and communications technologies is therefore of paramount importance, and the characterization of the scenario in which these technologies need to operate is fundamental to obtain meaningful results. In this paper, we present the implementation of three reference \gls{psc} scenarios, which are open source and made publicly available to the research community, describing the incidents, the mobility and applications of first responders, and providing examples on how a mmWave-based \gls{ran} can support high-traffic use cases. Moreover, we present the implementation of two novel mobility models for ns-3, which can be used to enable the simulation of realistic \gls{psc} scenarios in ns-3.

\end{abstract}


\maketitle

\section{Introduction}\label{sec:intro}
\glsresetall

\begin{picture}(0,0)(0,-470)
\put(0,0){
\put(0,0){\small This paper has been submitted to WNS3 2019. Copyright may be transferred without notice.}}
\end{picture}

Public safety operations ensure a prompt and effective response during incident or disaster scenarios. The situations in which first responders operate are often complex, involve multiple agencies and stakeholders who require coordination, and primary resources such as stable power supply may not be available. In this context, as discussed in~\cite{baldini2014survey}, the usage of information and communication technologies is seen as a promising enabler of improved response strategies during public safety emergencies. In particular, the next generation of \glspl{psc} is expected to provide first responders (i) enhanced coordination, for example by allowing the \gls{ic} station to experience through video what the responders in the field are witnessing; (ii) augmented awareness of the environment in which the operators engage, through sensing and localization; and (iii) enhanced response capabilities, thanks to the possibility of remotely controlling robots and drones~\cite{mezzavilla2018public}.

The design of \gls{psc} technologies has therefore been an important research topic for years, with studies that resulted in dedicated \gls{psc} standards~\cite{survey3}, or in the possibility of re-using commercial wireless technologies and networks to assist first responders~\cite{lin2014overview,mezzavilla2018public}. The suitability of a certain radio technology for \gls{psc}, however, largely depends on the context in which the communication is needed~\cite{theofanos2017usability}, and the definition of the incident scenario plays an important part in the characterization of this context. For example, it may be infeasible to use a fixed infrastructure during earthquakes or wildfire events, thus prompting the deployment of ad hoc networks~\cite{merwaday2015uav}, while other events (e.g., car accidents, shootings) may happen in areas served by dedicated or commercial fixed infrastructure, which can consequently be exploited by the first responders.

The performance evaluation of the different communication technologies must therefore be tailored to the specific public safety scenario under investigation. Consequently, a detailed and meaningful definition of the incident or disaster scenario is the first step of any performance assessment campaign, as also claimed in~\cite{choong2017incident}. Moreover, it can be useful to define reference scenarios, which can be shared and re-used for the evaluation of different communications technologies, so that it is possible to (i) directly compare operations with different protocol stacks and network architectures, and (ii) test new technologies, such as mmWave communications, currently not considered in highly demanding public safety scenarios~\cite{mezzavilla2018public}.

In this paper, we focus on the simulation of medium-scale scenarios using the popular open source network simulator ns-3, which is capable of modeling a wide range of wireless protocol stacks~\cite{lena} and networking protocols~\cite{CASONI201681}. In this regard, we will discuss and review the elements that should be modeled and considered when specifying public safety scenarios in ns-3. We will also present the implementation of three possible reference scenarios, derived from publicly available after-actions reports or public safety incident libraries~\cite{choong2017incident}: a multi-vehicle accident, involving a fuel truck, a chemical plant explosion and a high-school shooting. For each scenario, we will discuss the relevant modeling choices, and present some example results based on a mmWave cellular stack. The code for the scenarios is open source and publicly available,\footnote{The repository can be found at \url{https://github.com/signetlabdei/mmwave-psc-scenarios}} so that the research community can test the performance of other wireless networking solutions as well. The final contribution of this paper is the implementation of two new mobility models for ns-3, which are particularly useful for the modeling of public safety scenarios with obstacles and group mobility, like those we will present.

The rest of the paper is organized as follows. In Section~\ref{sec:psc} we provide an overview of the state of the art on the research of next generation networks for \gls{psc} use cases. Then, in Section~\ref{sec:implementation} we describe the three reference scenarios, together with examples on the performance of a communication infrastructure based on mmWave networks. In Section~\ref{sec:mobModels} we present the new mobility models we implemented, and finally we conclude the paper and provide suggestions for future work in Section~\ref{sec:conclusions}.

\section{Next Generation Public Safety Communications}
\label{sec:psc}
As mentioned in Sec.~\ref{sec:intro}, the research community is identifying the main trends in wireless communications technologies that can be applied to public safety use cases and scenarios. An exhaustive discussion can be found in~\cite{survey3}.

A first tendency is related to the study and assessment of the use in \gls{psc} of commercial devices, implementing the LTE protocol stack, and to the deployment of dedicated public safety networks based on LTE~\cite{doumi2013lte}. The LTE mobile broadband technology is indeed seen as a first step to increase the network capacity available to the first responders. The LTE specifications in Release 12 and 13 have also been enhanced to account for specific needs of \gls{psc}, such as off-coverage device-to-device and group communications. Moreover, in the United States, there has been a recent launch of a nation-wide LTE network for public safety operators (i.e., FirstNet~\cite{desourdis2015building}), in which a 20 MHz chunk of spectrum in the 700 MHz band has been prioritized for \gls{psc}. The goal of FirstNet is to promote interoperability and high performance networking, while reducing deployment and operational costs thanks to the re-use of commercial devices and specifications.
In ns-3, it is possible to study these scenarios using the LTE module~\cite{lena}, which has recently been extended to also model device-to-device operations~\cite{Rouil:2017:IVL:3067665.3067668}.

A second direction studies the feasibility of 5G and beyond technologies in the domain of public safety communications. For example, \gls{uav}-based communications, which have recently started being investigated by the 3GPP~\cite{lin2018sky}, are considered as natural candidates to provide coverage extension in challenging scenarios and assist during events in which the fixed infrastructure is not available~\cite{merwaday2015uav,naqvi2018drone}.
Another next-generation \gls{psc} enabler that has recently gained the attention of the networking researchers is the communication at mmWave frequencies. MmWaves are one of the main novelties of the fifth generation (5G) of cellular networks~\cite{38300}, and enable ultra-high data rate communications, thanks to the wide availability of bandwidth. Communications at such high frequencies, however, present a number of challenges that need to be addressed before mmWaves can be used for \gls{psc}, given the high reliability requirement during critical operations~\cite{RanRapE:14}. In this regard, the high propagation loss limits the communication range of devices operating at mmWave frequencies, which can be however increased using directional transmissions through beamforming. Moreover, mmWave frequencies suffer blockage from common materials, such as brick and mortar, and from the human body as well. Reference~\cite{mezzavilla2018public} discusses the potentials and challenges of mmWaves for \gls{psc}, with a use case in which a \gls{uav} uses a high-frequency link to transmit monitoring information (e.g., video from multiple cameras and sensors) from a wildfire scenario to a remote \gls{ic} stations. MmWave-based scenarios can also be studied using ns-3, using the contributed code of the cellular mmWave module~\cite{mezzavilla2017end} or of the 802.11ad module~\cite{Assasa:2016:IEW:2915371.2915377}.

Despite the numerous studies related to mmWave communications in cellular networks, their suitability for \gls{psc} scenarios is still to be defined~\cite{mezzavilla2018public}. In this regard, and, in general, to understand how next-generation technologies perform for \gls{psc}, it is important to define compelling scenarios, with a realistic deployment of the first responders, of the obstacles in the scenario and of the communication infrastructure, and a proper modeling of the interactions between first responders and the involved parties. In the following section, we will present the implementation of three public safety scenarios for ns-3, in which the communication infrastructure targets a mmWave deployment, but that can be adapted and used to study any wireless \gls{psc} use case.

\section{Implementation of Public Safety Scenarios}
\label{sec:implementation}

The three public safety scenarios we implemented feature a multi-vehicle accident~\cite{mva}, a chemical plant explosion~\cite{ogata2016identifying} and a public school shooting incident~\cite{shooting}.
The definition of each of these \gls{psc} scenarios involves a two-step process. Indeed, it is first necessary to define the general elements related to the incident, which are independent of the communication technology being tested (e.g., the timeline of the events, the operations that the first responders need to perform). Next, it is possible to characterize the communication infrastructure whose performance needs to be evaluated.

In the first phase of the scenario modeling, we studied the after-actions reports and the scenarios proposed in~\cite{mva,ogata2016identifying,shooting}, to elaborate an accurate description of the incidents, which were then implemented in ns-3 and will be described in the following paragraphs. In particular, the details on which we focused are (i) the incident area, for which we specified the dimensions, and the possible obstacles, buildings and obstructions; (ii) the first responders, indicating how many operators are deployed on the scene, with which kind of equipment, and with which mobility patterns they move in the incident area; and (iii) the communication requirements, in terms of applications needed by the public safety personnel (e.g., video monitoring, data collection from sensors, sharing of location information).
Then, for the characterization of the communication technologies and networking architectures, it is important to define the nature of the deployment (e.g., ad hoc, based on shared or dedicated infrastructure), the protocol stack that will be used on the devices and which kinds of \gls{ran} and core network will be considered. Moreover, the specifications should include also the source model for the applications that the first responders use over the network.

The three scenario that we developed and present in this paper are publicly available, as mentioned in Sec.~\ref{sec:intro}. In the same repository, we also provide the \texttt{install.sh} script, which checks and installs the dependencies that are needed to run the examples. In this case, given that the communication technology that is simulated is based on mmWaves, the script clones, configures, patches and builds the ns-3 forks with the mmWave cellular network implementation described in~\cite{mezzavilla2017end}, or on the mmWave \gls{iab} extension illustrated in~\cite{polese2018end}. A possible integration in the ns-3 App Store will be considered.

Finally, in order to avoid unneeded code duplication for the setup of the public safety simulation scenarios, we implemented a helper, with static methods, in the \texttt{ps-simulation-config} files. This helper can be used, for example, to set up the core network and the connectivity to the public internet in every scenario, with the \texttt{CreateInternet} and \texttt{InstallUeInternet} methods, to deploy the nodes, buildings and obstacles for the scenario, and to set the path for the log traces.

\subsection{Multi-vehicle Accident} 
\label{sub:multi_vehicle_accident}
The multi-vehicle accident scenario we implemented is described in~\cite{mva} as a possible future public safety technology scenario. An incident involving multiple vehicles is reported, but with limited information on the location and number/kind of vehicles involved. A fire chief arrives at the scene and assesses the precise location and the nature of the incident. It involves three vehicles: an overturned liquid propane truck, and two passenger vehicles badly damaged, with injured victims. After the initial assessment, firefighters, emergency services, the police and electric utility personnel are notified and reach the area. In the implemented scenario, we focus on the time interval during which the firefighters arrive at the scene and begin the safety operations, in order to secure the vehicles and avoid a possible fire or explosion of the liquid propane truck. We consider $N$ first responders on the ground, each equipped with a body-worn camera that streams to the public safety network and the \gls{ic} station the scene from the point of view of the operator. Moreover, some of the firefighters ($d N$, with $0 \le d \le 1$) also wear augmented reality displays which stream the view of the other responders. The scenario could be further extended by adding other traffic sources and sinks, but, for this preliminary mmWave-based evaluation, we focus on a video application which should be enough stress to the available resources.

\begin{figure}[t]
	\centering
	\includegraphics[width=.65\columnwidth]{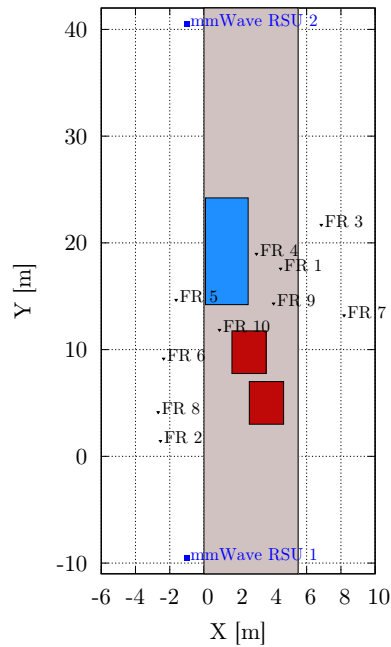}
	\caption{Multi-vehicle accident scenario, with the random initial positions of the first responders, the mmWave \glspl{rsu}, the cars (in red) and the truck (in blue).}
	\label{fig:mva}
\end{figure}

The scenario is implemented in the \texttt{mva-scenario.cc} script, and can be configured with respect to a number of parameters (e.g., the number of first responders $N$, the rate of the video application, the number of trucks and cars involved in the accident). The default incident area is illustrated in Figure~\ref{fig:mva}. In particular, we consider a section of the road with a width of 5.5 m, and a length of 25 m for the incident area. The two cars and the truck are randomly placed in the street, and are implemented using the \texttt{buildings} module of ns-3. The method \texttt{CreateRandomObstacle} of the helper class is used to deploy each obstacle, and accepts as input parameters the width and length of the street and of the obstacles, which can be either aligned to the direction of the street (as shown in Figure~\ref{fig:mva}), or orthogonal to it. The first responders are also dropped in random locations, using the \texttt{DropFirstResponders} method of the helper. The latter relies on ns-3's \texttt{OutdoorPositionAllocator} for the initial position allocation, and assigns to the first responder nodes a random walk mobility model which prevents them from crossing the obstacles. More details on the mobility model will be provided in Sec.~\ref{sec:mobModels}. Finally, the video streaming applications are modeled in terms of a constant bitrate flow on UDP, as in~\cite{mezzavilla2018public}. The method \texttt{SetupUdpFlow} of the helper takes care of configuring the endpoints of the communication, for both uplink and downlink streams, with randomized starting points for each flow.

In terms of communication infrastructure, in the example we provide we consider a mmWave \gls{ran}~\cite{mezzavilla2017end}, but the code for the scenario can be easily extended to also account for other kinds of wireless technologies available in ns-3 (e.g., LTE, Wi-Fi). In this case, two mmWave \glspl{rsu} are randomly placed along the incident scenario, at a default inter site distance of 50 m, in order to model a next-generation network deployment in which mmWave base stations are used for cellular and vehicle-to-infrastructure communications~\cite{giordani2017mmwave}. During the their intervention, the first responders can tap into the resources provided by the \glspl{rsu} in order to support high-bitrate video streaming. Moreover, in order to enhance the communication reliability, a \gls{lte} base station is also considered to carry first responders data traffic, and is deployed at a distance of 500 m from the scenario, in order to model an urban configuration of a FirtsNet-like network. The first responder equipment can access the network using multi connectivity over the different links, following the approach proposed in~\cite{poleseHo}.

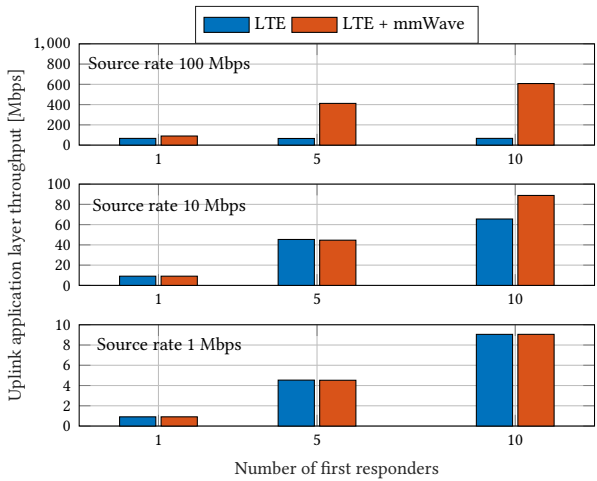
\begin{figure}[t]
	\centering
	\setlength\fwidth{.85\columnwidth}
	\setlength\fheight{.6\columnwidth}
%
%
\definecolor{mycolor1}{rgb}{0.00000,0.44700,0.74100}%
\definecolor{mycolor4}{rgb}{0.6784,0.8471,0.9020}%
\definecolor{mycolor2}{rgb}{0.85000,0.32500,0.09800}%
\definecolor{mycolor5}{rgb}{1.0000,0.6275,0.4784}%
\definecolor{mycolor3}{rgb}{0.92900,0.69400,0.12500}%
\definecolor{mycolor6}{rgb}{1.0000,0.9804,0.8039}%
\begin{tikzpicture}
\pgfplotsset{every tick label/.append style={font=\scriptsize}}

\begin{axis}[%
width=0.951\fwidth,
height=0.265\fheight,
at={(0\fwidth,0.735\fheight)},
scale only axis,
bar shift auto,
xmin=-1,
xmax=12,
xtick={ 1,  5, 10},
xlabel style={font=\color{white!15!black}},
ymin=0,
ymax=1000,
xmajorgrids,
ymajorgrids,
ylabel style={font=\color{white!15!black}},
axis background/.style={fill=white},
legend style={legend cell align=left, font=\footnotesize, at={(0.5, 1)}, anchor=south, align=left, draw=white!15!black},
legend columns=2
]
\addplot[ybar, bar width=0.914, fill=mycolor1, draw=black, area legend] table[row sep=crcr] {%
1	66.3962427474747\\
5	65.5961788062443\\
10	66.5554676951331\\
};
\addlegendentry{LTE}

\addplot[ybar, bar width=0.914, fill=mycolor2, draw=black, area legend] table[row sep=crcr] {%
1	89.9101877979798\\
5	411.905231808999\\
10	607.956845624313\\
};
\addlegendentry{LTE + mmWave}

\end{axis}

\begin{axis}[%
width=0.951\fwidth,
height=0.265\fheight,
at={(0\fwidth,0.368\fheight)},
scale only axis,
bar shift auto,
xmin=-1,
xmax=12,
xtick={ 1,  5, 10},
xlabel style={font=\color{white!15!black}},
ymin=0,
ymax=100,
xmajorgrids,
ymajorgrids,
ylabel shift=2pt,
ylabel style={font=\footnotesize\color{white!15!black}},
ylabel={Uplink application layer throughput [Mbps]},
axis background/.style={fill=white},
]
\addplot[ybar, bar width=0.914, fill=mycolor1, draw=black, area legend] table[row sep=crcr] {%
1	9.09394747474748\\
5	45.3441492745638\\
10	65.5473522075776\\
};

\addplot[ybar, bar width=0.914, fill=mycolor2, draw=black, area legend] table[row sep=crcr] {%
1	9.0996405010101\\
5	44.6419617263545\\
10	88.8080767958588\\
};

\end{axis}

\begin{axis}[%
width=0.951\fwidth,
height=0.265\fheight,
at={(0\fwidth,0\fheight)},
scale only axis,
bar shift auto,
xmin=-1,
xmax=12,
xtick={ 1,  5, 10},
xlabel style={font=\footnotesize\color{white!15!black}},
xlabel={Number of first responders},
ymin=0,
ymax=10,
xmajorgrids,
ymajorgrids,
ylabel style={font=\color{white!15!black}},
axis background/.style={fill=white},
]
\addplot[ybar, bar width=0.914, fill=mycolor1, draw=black, area legend] table[row sep=crcr] {%
1	0.909652598473875\\
5	4.53481707318952\\
10	9.05190393693028\\
};


\addplot[ybar, bar width=0.914, fill=mycolor2, draw=black, area legend] table[row sep=crcr] {%
1	0.910590726539779\\
5	4.52135963269054\\
10	9.05438383838384\\
};

\end{axis}

\node[anchor=north,xshift=0pt,yshift=-20pt] at (1.2, 2) {\footnotesize Source rate 1 Mbps};
\node[anchor=north,xshift=0pt,yshift=-20pt] at (1.2, 3.85) {\footnotesize Source rate 10 Mbps};
\node[anchor=north,xshift=0pt,yshift=-20pt] at (1.2, 5.73) {\footnotesize Source rate 100 Mbps};

\end{tikzpicture}%
	\caption{Aggregate uplink throughput for the video streaming in the multi-vehicle accident scenario, for different video source rates and different numbers of first responders.}
	\label{fig:mva-results}
\end{figure}

A set of example results is reported in Figure~\ref{fig:mva-results}, where we compare the usage in this scenario of a 4G communication infrastructure, based on the \gls{lte} base station only, and the 5G setup previously described. The metric we consider is the application layer throughput, i.e., the sum of the video throughput received at the \gls{ic} station and generated by the uplink feed from the first responders. We vary the source rate of the video stream (i.e., 1, 10 or 100 Mbps), and the number of first responders actively transmitting uplink video on the scene (i.e., 1, 5 or 10). As it can be seen, the \gls{lte} connection is enough for use cases with a low source rate, or few first responders. As the traffic increases, however, the \gls{lte} base station saturates its capacity, and the solution with a combined mmWave and \gls{lte} usage dramatically improves the overall performance. This example showcases one of the main benefits of a mmWave-based network for \gls{psc}, i.e., the high capacity that this kind of \gls{ran} can offer.


\subsection{Chemical Plant Explosion} 
\label{sub:chemical_plant_explosion}

\begin{figure}[t]
	\centering
	\includegraphics[width=.95\columnwidth]{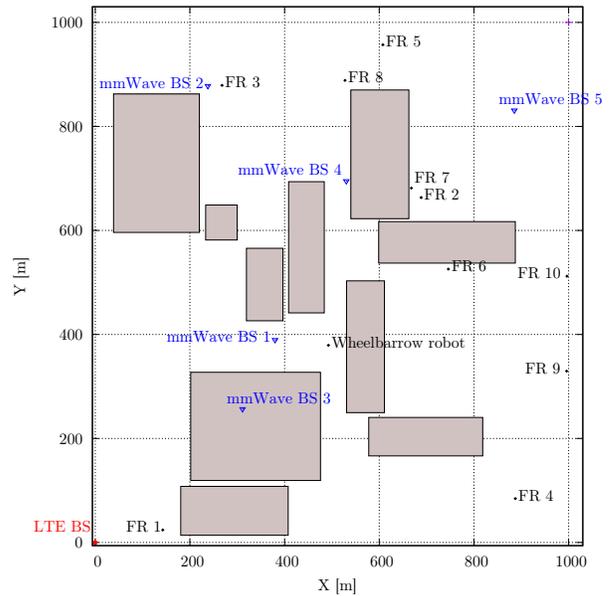}
	\caption{Chemical plant explosion scenario. The random deployment features 10 buildings (the gray rectangles), 10 first responders and a wheelbarrow robot. Communication is provided by the fixed infrastructure of the plant, with five mmWave and one LTE base stations.}
	\label{fig:chem}
\end{figure}

The chemical plant explosion is inspired by the incident briefly described in~\cite{ogata2016identifying}. A large explosions takes place at a chemical plant. There exists the possibility of a hazardous chemical leak, as well as toxic smoke emission from the chemicals on fire. The incident is notified to the public safety authorities, and firefighters, police and emergency services are deployed in the area. In this implementation, we consider the time interval in which the first responders are already at the incident scene, and stream data from body-worn cameras and sensors to the \gls{ic} station, where a monitor shows the status of each first responder. Moreover, an explosive ordinance disposal team also operates in the area, using a wheelbarrow robot\footnote{\url{https://www.army-technology.com/projects/wheelbarrowmk9/}} to safely inspect and operate in the critical on-fire area, where the risk of chemical leaks and explosions creates a danger for human first responders. The robot is equipped with high-resolution cameras and is remotely controlled from the IC station.

Figure~\ref{fig:chem} illustrates a random realization of the scenario, implemented in the \texttt{chemical-plant-scenario.cc} script. As for the other examples, it is possible to configure from the command line a number of parameters, e.g., the number of first responders, the data rate for video and control applications, and the size of the chemical plant area. With the default settings, shown in Figure~\ref{fig:chem}, the chemical plant covers an area of 1 km$^2$, and hosts 10 buildings randomly deployed using the \texttt{CreateRandomBuildings} method of the helper. Both the first responders and the wheelbarrow robot use the building-aware mobility model described in Sec.~\ref{sec:mobModels}. The robot operates in a constrained area at the center of the scenario, where the explosion happened, while the first responders are free to roam in the whole chemical plant area.

For the communications, in this case we envision a next-generation network scenario in which the first responders re-use the private networking infrastructure of the chemical plant, thanks to the coordination of the plant-managed core network and the public safety core infrastructure for authentication purposes. We consider $N_{bs} = 5$ mmWave base stations, that usually provide cellular connectivity to the operators of the chemical plant and to the surveillance and monitoring equipment. Moreover, we compare the mmWave-only scenario with a configuration in which an \gls{lte} base station is also deployed in the area, and used to provide a reliable coverage layer to the wheelbarrow robot and prioritize the data flow for its remote control.

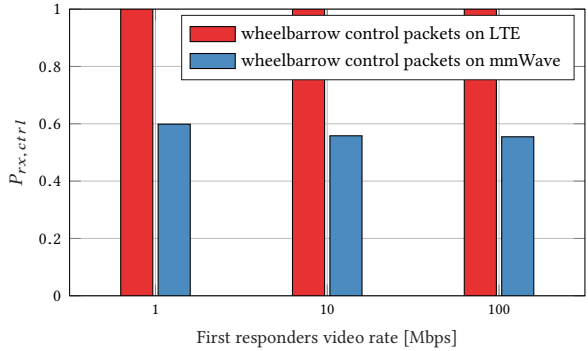
\begin{figure}[t]
	\centering
	\setlength\fwidth{.85\columnwidth}
	\setlength\fheight{.45\columnwidth}
%
\definecolor{mycolor1}{rgb}{0.90471,0.19176,0.19882}%
\definecolor{mycolor2}{rgb}{0.29412,0.54471,0.74941}%
\definecolor{mycolor3}{rgb}{0.37176,0.71765,0.36118}%
\begin{tikzpicture}
\pgfplotsset{every tick label/.append style={font=\scriptsize}}

\begin{axis}[%
width=0.951\fwidth,
height=\fheight,
at={(0\fwidth,0\fheight)},
scale only axis,
bar shift auto,
xmin=0.5,
xmax=3.5,
xtick={1,2,3},
xticklabels={1, 10, 100},
xlabel style={font=\footnotesize\color{white!15!black}},
xlabel={First responders video rate [Mbps]},
ymin=0,
ymax=1,
ylabel style={font=\footnotesize\color{white!15!black}},
ylabel={$P_{rx, ctrl}$},
axis background/.style={fill=white},
title style={font=\bfseries},
ylabel shift=-2pt,
xmajorgrids,
ymajorgrids,
legend style={font=\footnotesize, legend cell align=left, align=left, draw=white!15!black}
]
\addplot[ybar, bar width=0.187, fill=mycolor1, draw=black, area legend] table[row sep=crcr] {%
1	1\\
2	0.99993031358885\\
3	1\\
};
\addlegendentry{wheelbarrow control packets on LTE}

\addplot[ybar, bar width=0.187, fill=mycolor2, draw=black, area legend] table[row sep=crcr] {%
1	0.598621420996819\\
2	0.558118466898955\\
3	0.554660278745645\\
};
\addlegendentry{wheelbarrow control packets on mmWave}

\end{axis}
\end{tikzpicture}%
	\caption{Probability $P_{rx, ctrl}$ of receiving a control packet for the wheelbarrow robot in the chemical plant scenario, when using a dedicated \gls{lte} link or a mmWave connection shared with the video streaming feed of the first responders and the wheelbarrow robot itself. Different source rates for the single video feed are tested.}
	\label{fig:chem-results}
\end{figure}

Figure~\ref{fig:chem-results} reports an example of the results that can be obtained in this scenario, comparing the operations with the dedicated \gls{lte} link and with a shared mmWave connection for the wheelbarrow robot control. The metric we report is the probability $P_{rx, ctrl}$ of receiving a control packet, defined as the ratio between the received and transmitted control packets. As it can be seen, the mmWave \gls{ran} does not provide enough reliability to the control link, which exhibits a high loss even if the video traffic generated by the wheelbarrow robot itself and the other first responders is low (i.e., 1 Mbps). Moreover, as the rate of the video source increases, $P_{rx, ctrl}$ decreases, showing that, without prioritization and a proper management of the quality of service of the different end-to-end connections, the resource-consuming video traffic affects the performance of the wheelbarrow control commands (which have a lower rate of 500 kbps). This example calls for the study of more refined deployment strategies, with multi connectivity and, where needed, ad hoc mmWave relay nodes to provide a better coverage in \gls{psc} scenarios, and of service differentiation and quality of service management at mmWave frequencies.

\subsection{Public School Shooting} 
\label{sub:public_school_shooting}

The last scenario is a public school shooting, described in~\cite{shooting}. The original settings foresee a large scale scenario, with approximately 2000 students and 110 first responders that arrive at the public school after an active shooter barricades into a classroom. The public safety operators include the officers responsible for securing the perimeter of the area, an \gls{ic} unit, a SWAT strike team, and the emergency services and logistics operators. The time line of the events spans 4 hours, from the arrival on the scene of the first enforcement officers, to the operations of the SWAT team and the securing of the area. However, given the simulation complexity that a mmWave-based stack often yields, we focus in our implementation on the time interval in which the SWAT team moves inside the school building to reach the shooter's position. We consider 4 SWAT teams, with 4 officers each, moving from the 4 corners of the building towards its center, where the shooter is located. As in the multi-vehicle accident scenario, the officers of the strike team wear a helmet camera that streams a high quality video towards the \gls{ic} station, which monitors the development of the operation.

\begin{figure}[t]
	\centering
	\includegraphics[width=.95\columnwidth]{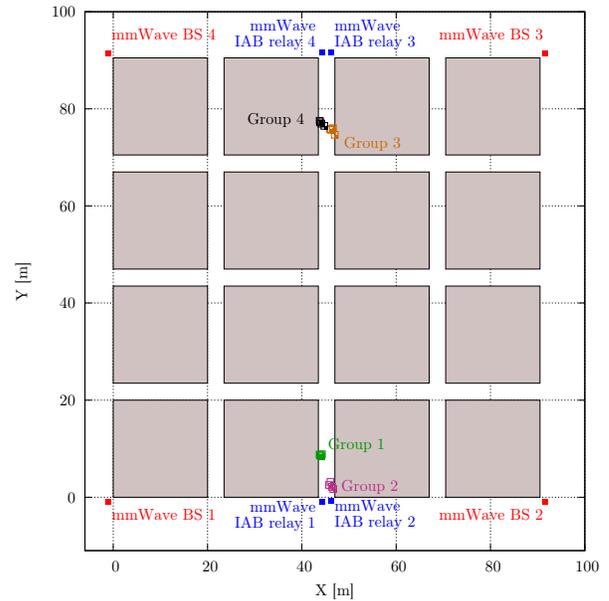}
	\caption{Public school shooting scenario. The deployment is indoors, in a building with 16 rooms and corridors in between. The communications are provided by the four mmWave base stations in the corners, and, where needed, by the mmWave \gls{iab} relays.}
	\label{fig:shoot}
\end{figure}

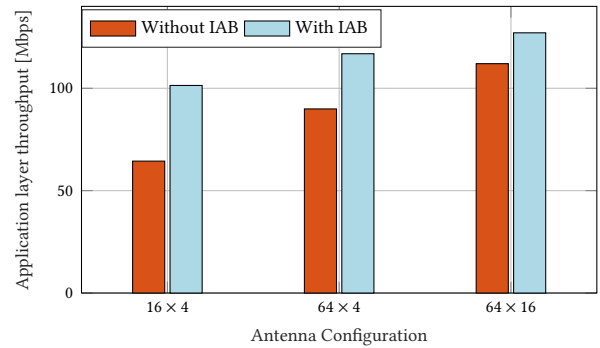
\begin{figure}[t]
	\centering
	\setlength\fwidth{.85\columnwidth}
	\setlength\fheight{.45\columnwidth}
%
%
\definecolor{mycolor6}{rgb}{0.00000,0.44700,0.74100}%
\definecolor{mycolor2}{rgb}{0.6784,0.8471,0.9020}%
\definecolor{mycolor1}{rgb}{0.85000,0.32500,0.09800}%
%
\begin{tikzpicture}
\pgfplotsset{every tick label/.append style={font=\scriptsize}}

\begin{axis}[%
width=0.951\fwidth,
height=\fheight,
at={(0\fwidth,0\fheight)},
scale only axis,
bar shift auto,
xmin=0.5,
xmax=3.5,
xlabel style={font=\footnotesize\color{white!15!black}},
xlabel={Antenna Configuration},
xtick={1,2,3},
xticklabels={$16 \times 4$, $64 \times 4$, $64 \times 16$},
ymin=0,
ymax=140,
ylabel style={font=\footnotesize\color{white!15!black}},
ylabel={Application layer throughput [Mbps]},
axis background/.style={fill=white},
title style={font=\bfseries},
xmajorgrids,
ymajorgrids,
legend columns=2,
legend style={at={(0, .99)}, anchor=north west, font=\footnotesize,legend cell align=left, align=left, draw=white!15!black}
]
\addplot[ybar, bar width=0.187, fill=mycolor1, draw=black, area legend] table[row sep=crcr] {%
1	64.425984\\
2	89.9143246353405\\
3	112.018698722358\\
};
\addlegendentry{Without IAB}

\addplot[ybar, bar width=0.187, fill=mycolor2, draw=black, area legend] table[row sep=crcr] {%
1	101.381586980392\\
2	116.87302172549\\
3	127.080211137255\\
};
\addlegendentry{With IAB}

\end{axis}
\end{tikzpicture}%
	\caption{Aggregate throughput for the public school shooting scenario. The x axis reports different antenna configurations, where the first value is the number of antennas at the mmWave base stations and \gls{iab} relays, and the second that in the first responder devices.}
	\label{fig:swat-results}
\end{figure}

The code for this scenario is in the \texttt{psc-shooting-swat.cc} script, and an example of the deployment is shown in Figure~\ref{fig:shoot}. Contrary to the two other scenarios described in this paper, this is an indoor scenario, where the nodes are inside a building with a square grid of four rooms on each side. Each square room has a side of 20 meters. The four SWAT teams move in a coordinated way, with the officers following the leader of the team using the group mobility model described in Sec.~\ref{sec:mobModels}.

In this scenario, the main challenge for the deployment of an ad hoc mmWave infrastructure to assist the first responders is the lack of \gls{los} given by the presence of rooms and walls in the scenario. We consider four mmWave base stations with a wired backhaul deployed at the corners of the building, i.e., where the SWAT teams enter the building. Thus, in order to provide assistance as the officers move through the scenario, we investigate a solution with nomadic relays, which move along with one of the operators and are positioned in areas which guarantee \gls{los} to both the officers and the base stations with a wired connection to the core network. The relays use the integrated access and backhaul stack recently studied by the 3GPP~\cite{polese2018end}.

Figure~\ref{fig:swat-results} compares the performance in this scenario of solutions with and without \gls{iab}, when varying the number of antennas at the base stations (16 or 64) and at the first responder devices (4 or 16). The metric that is reported in Figure~\ref{fig:swat-results} is the total throughput of the data received at the \gls{ic} station, as for the multi-vehicle accident scenario. It can be seen that the architecture with nomadic \gls{iab} nodes always delivers a better performance compared to that without relays, in which the users only exploit reflections to receive the signals. Moreover, the performance gap increases as fewer antenna elements are used. These results confirm that relay-based ad hoc scenarios are useful to increase the coverage in mmWave \gls{psc} scenarios, and call for additional studies on how to efficiently deploy \gls{iab} networks for \gls{psc}.


\section{New Mobility Models to Enable PSC Scenarios}
\label{sec:mobModels}

In order to support the \gls{psc} scenarios described in Sec.~\ref{sec:implementation}, and, in general, to increase the modeling capabilities of the \texttt{mobility} and \texttt{buildings} modules of ns-3, we implemented two new mobility models that will be described in the following paragraphs.

\subsection{Building-Aware Random Walk} 
\label{sub:building_aware_random_walk}
The first model is an adaptation of the \texttt{RandomWalk2dMobility\-Model} that also takes into account the presence of \texttt{Building} objects in the simulation scenario and avoids them, so that the node is always outdoors. This model can be useful in cases in which the \texttt{Building} objects simulate obstacles, which the node is not supposed to go through, or outdoor-only scenarios, in which the nodes do not need to operate inside the actual buildings. Moreover, while this implementation specifically refers to a random walk mobility model, the same approach can be easily applied to make other ns-3 mobility models avoid buildings.

The implementation can be found in the \texttt{RandomWalk2dOutdoor\-MobilityModel} class, which is part of the \texttt{buildings} module. It maintains a logic similar to that of \texttt{RandomWalk2dMobility\-Model}, with the  \texttt{DoInitializePrivate} and \texttt{DoWalk} methods that are periodically called to select a random direction and velocity and walk along them, respectively. Both the standard and the building-aware implementations make it possible to specify a bounding area in which the nodes mobility is constrained, through a \texttt{Rectangle} attribute. When a node reaches a border of the bounding box, it rebounds with a simmetric direction and velocity.

Moreover, the building-aware implementation performs some additional checks to avoid intersecting a building in the \texttt{DoWalk} method. First, the next expected position is extracted, based on the current velocity and direction. Then, the method \texttt{IsLineClearOf\-Buildings} draws an imaginary line between the current and the next position, and check if the line intersects any object from the \texttt{BuildingList}, applying the separating axis test also used in the propagation classes of the \texttt{mmwave} module~\cite{zhang20173gpp}. If the line is clear of buildings, then the walk proceeds as expected. Otherwise, the walk continues until the intersection with the intersecting \texttt{Building} object is reached. Then, a new course is extracted in the \texttt{AvoidBuilding} method, ensuring that it will not intersect other buildings until the next update.

\begin{figure}[t]
	\centering
	\includegraphics[width=\columnwidth]{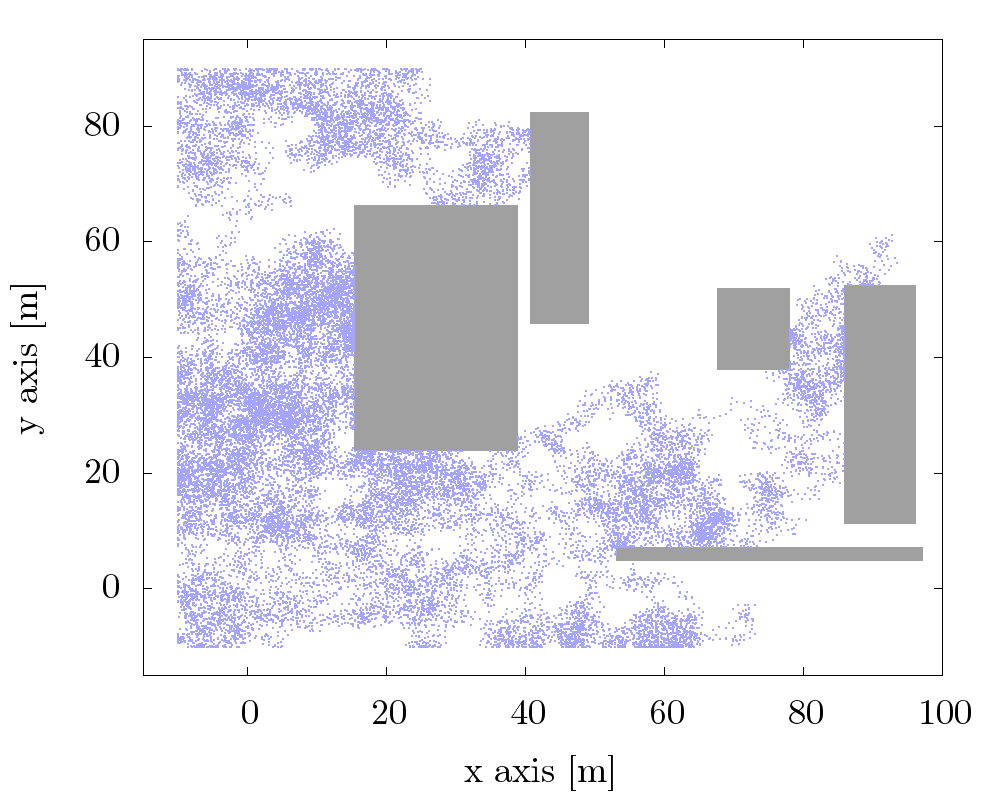}
	\caption{Example of building-aware random walk. The blue dots represent the visited positions, and the gray rectangles are \texttt{Building} objects.}
	\label{fig:outdoor}
\end{figure}

As an example, we report in Figure~\ref{fig:outdoor} a realization of a random walk generated using the \texttt{RandomWalk2dOutdoor\-MobilityModel} class. The testing scenario features 5 randomly generated obstacles in a rectangular bounding box between coordinates $(-10, -10)$ m and $(100, 90)$ m.


\subsection{Master-Slave Mobility Model} 
\label{sub:master_slave_mobility_model}
The second contribution is a set of mobility models in which a group of nodes (i.e., the \textit{slaves}) follow the mobility of a \textit{master} node. The implementation follows the idea of the \gls{rpgm} model presented in~\cite{hong1999group,camp2002survey}: the master node acts as a group mobility vector, and the other nodes follow its mobility by sampling random positions and directions around those of the master.

\begin{figure}[t]
	\centering
	\includegraphics[width=\columnwidth]{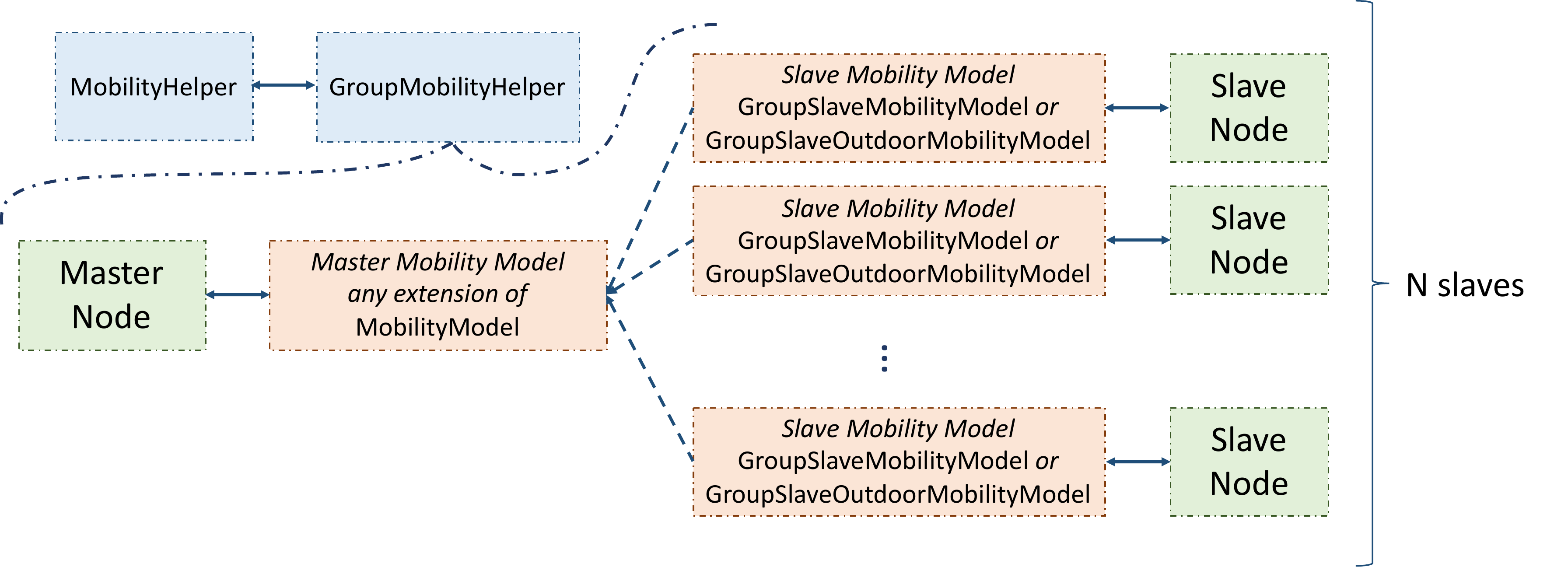}
	\caption{Visual representation of master-slave mobility model code structure.}
	\label{fig:ms}
\end{figure}

The structure of the implementation can be found in Figure~\ref{fig:ms}.
The core is represented by the \texttt{GroupSlaveMobilityModel} class, which extends the \texttt{MobilityModel} class. In order to make the implementation as modular as possible, each \texttt{GroupSlaveMobilityModel} associated with a certain master node holds a pointer to the object representing the \texttt{MobilityModel} of the master. In this way, the master can use any implementation of the mobility models available in ns-3. According to the methodology described in~\cite{hong1999group}, the master mobility model is used as a reference point, and, when the \texttt{GetPosition} method is called for the \texttt{GroupSlaveMobilityModel} objects, a random position around the reference point is returned. The randomness of the position can be controlled through an ns-3 random variable object, which can be configured through the \texttt{RandomVariable} attribute of the \texttt{GroupSlaveMobilityModel} class. The default is a Gaussian random variable, with mean $\mu = 0$ m, standard deviation $\sigma = 1$ m and bounded at 20 m. Moreover, in order to track the evolution of the reference point mobility in the scenario, the method \texttt{MasterCourseChanged} of the \texttt{GroupSlaveMobilityModel} objects is bound to the \texttt{CourseChange} callback of the master model, and triggers the same callback for the slave model.

The \texttt{GroupSlaveMobilityModel} class also features a protected virtual method \texttt{CheckForSpecialConditions}, that can be used to reject randomly sampled positions in special cases. In the default implementation, the method always returns true and any position is accepted. Nonetheless, it useful to extend this class to account, for example, for the presence of buildings in the scenario and reject positions that are inside them, as done in the \texttt{GroupSlaveOutdoorMobilityModel} class, which extends \texttt{Group\-SlaveMobilityModel} by overriding \texttt{CheckFor\-Special\-Con\-di\-tions}. This class is implemented in the \texttt{buildings} module, and is a natural complement of the \texttt{RandomWalk2dOutdoor\-Mobility\-Model} class described in Section~\ref{sub:building_aware_random_walk} in a master/slave context. Notice that, in general, it may not be possible to successfully return a slave position, for example because the master uses a non-building-aware mobility model and is inside a building, and the slave rejects indoor positions. While this eventuality should be avoided during the simulation scenario design, we added a limit to the maximum number of iterations for each sampled position, which can be tuned through the \texttt{MaxIterations} attribute. If the limit is reached, the simulation stops and the user is notified of the issue.

\begin{figure}
	\centering
	\begin{subfigure}[t]{\columnwidth}
		\centering
		\includegraphics[width=\columnwidth]{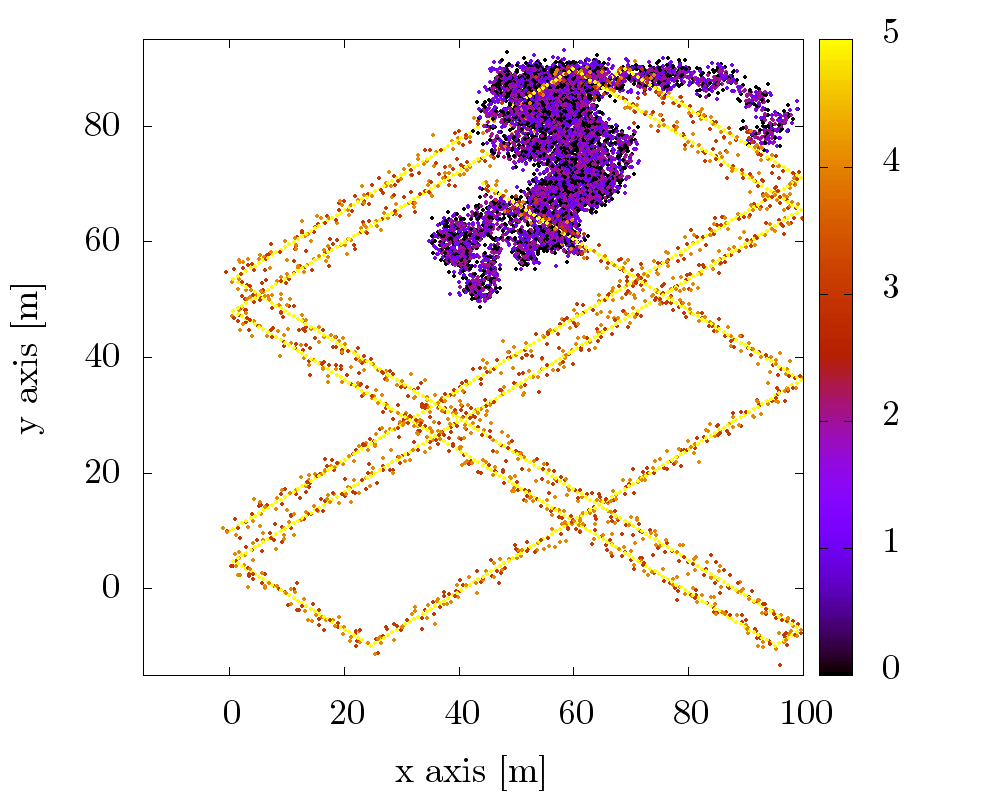}
		\caption{Example with two masters (node 2 moving with a random walk and 5 with a Gauss-Markov mobility model) with two slaves each, without buildings.}
		\label{fig:ms_any}
	\end{subfigure}
	\begin{subfigure}[t]{\columnwidth}
		\centering
		\includegraphics[width=\columnwidth]{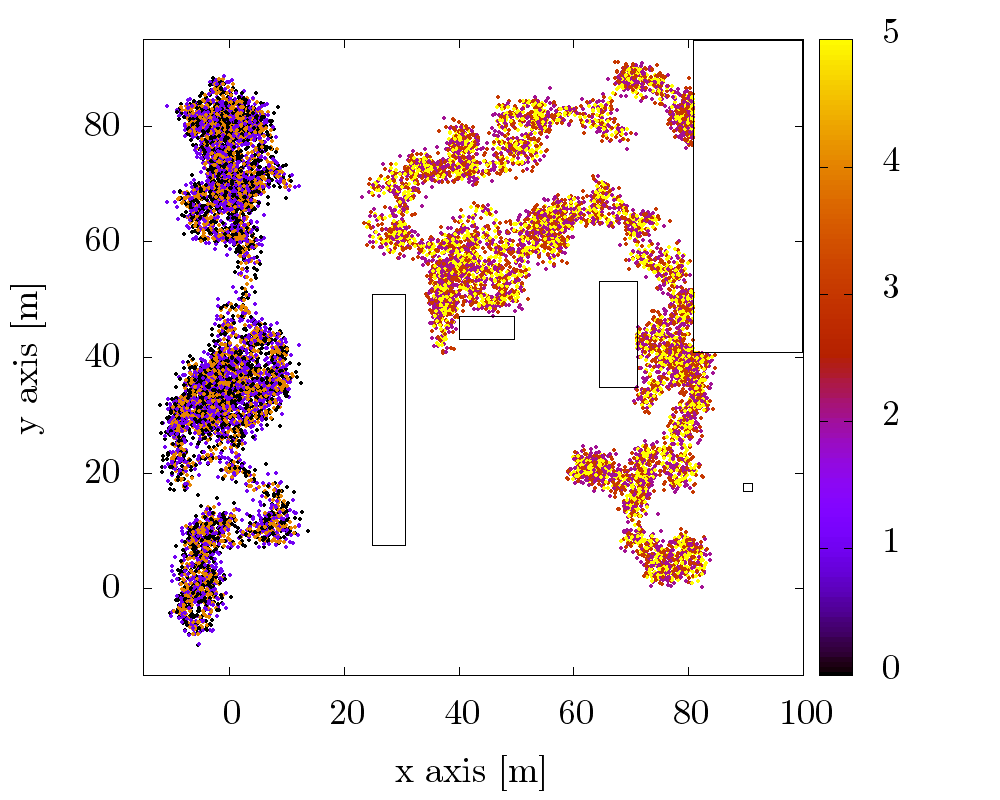}
		\caption{Example with two masters (both moving with a building-aware random walk mobility model) with two slaves each.}
		\label{fig:ms_bui}
	\end{subfigure}
	\caption{Examples of group mobility, with four slaves and two masters in scenarios with and without buildings. Each node has a different color, reported in the legend on the right. Nodes 0 and 1 are the slaves of node 2, while 3 and 4 of node 5.}
	\label{fig:example_ms}
\end{figure}

The setup is aided by a new helper, the \texttt{GroupMobilityHelper}, which takes care of configuring the relevant parameters in the master and slaves.
First, it is necessary to initialize a \texttt{MobilityHelper} for the master, which will handle the installation of the master mobility model in the master node. A pointer to this helper is stored in the \texttt{GroupMobilityHelper} object through the \texttt{SetMobilityHelper} method. Then, the \texttt{InstallGroupMobility} method of \texttt{Group\-Mobi\-lityHelper} can be called for a \texttt{NodeContainer} with the slave nodes. This method creates a master node, uses the stored mobility helper to install the master mobility model in the node, and finally installs and configures the slave mobility models. It also returns a \texttt{NodeContainer}, where the first node is the master, and the other nodes are the slaves it received as input. This set of operations can be repeated to install different master/slave mobility models on different sets of nodes. The \texttt{GroupMobilityHelper} has two attributes, i.e., two strings that accept the \texttt{TypeId} of the slave mobility model and of its random variable, in \texttt{GroupSlaveMobilityModel} and \texttt{PathDeviationRandomVariable} respectively.

Figure~\ref{fig:example_ms} reports examples with two different realizations of the mobility of 6 nodes, i.e., 2 masters with 2 slaves each. In particular, Figure~\ref{fig:ms_any} is a scenario without buildings, and we compare the behavior of three nodes moving with a Gauss-Markov mobility model (which correlates the next direction and velocity to the previous ones), and three moving according to a random walk, while Figure~\ref{fig:ms_bui} considers a deployment with 5 buildings or obstacles, and two groups of nodes using the \texttt{RandomWalk2dOutdoor\-Mobility\-Model}. In both figures it is possible to clearly identify two group behaviors, with the nodes with a common master moving with a similar pattern. As mentioned in Sec.~\ref{sub:public_school_shooting}, the master/slave mobility model can be used in scenarios where there is a group leader (e.g., a higher-ranking officer) and multiple members of the team who need to follow the leader, and the mobility pattern of the leader could be random and not pre-configurable in the simulation scenario~\cite{badia2007mobility}.


\section{Conclusions}
\label{sec:conclusions}
In this paper we presented the ns-3 implementation of three public safety scenarios (multi-vehicle accident, chemical plant explosion and high school shooting) that can be used as a reference for the evaluation of different protocol stacks and \gls{psc} technologies. In this regard, we briefly presented the main research trends related to the next generation of \gls{psc}, and illustrated that, for a preliminary assessment of their feasibility for public safety use cases, it is important to test their usage in realistic scenarios, in which the emergency or incident events are accurately modeled. After briefly reviewing the main characteristics that in general should be specified for the scenarios, we discussed the implementation details of each of them. Moreover, we presented some example results, based on a preliminary evaluation of a mmWave- and \gls{lte}-based deployment to support high-quality video streaming and reliable remote control. Finally, we presented the novel implementation of two mobility models that can be used to enhance the realism of the first responders movements during the simulation, i.e., for scenarios in which they need to avoid obstacles and in which some operators follow the actions of a common leader.

The implementation of the scenarios is publicly available, and can be reused by the \gls{psc} research community. As future work, we will expand the library of scenarios, and improve the variety and accuracy of the applications used by the first responders.

\vspace{-.1cm}
\begin{acks}
	This work was partially supported  by the U.S. Commerce Department/NIST through the project ``An End-to-End Research Platform for Public Safety Communications above 6 GHz''.
\end{acks}

\bibliographystyle{ACM-Reference-Format.bst}
\bibliography{bibl.bib}

\end{document}